\documentclass[twocolumn,amsmath,amssymb,aps,prl,groupedaddress,floatfix,showpacs,superscriptaddress]{revtex4-1}

\usepackage{graphicx}% Include figure files
\usepackage{dcolumn}% Align table columns on decimal point
\usepackage{bm}% bold math
\usepackage[usenames, dvipsnames]{color} 
\usepackage{braket} %for braket notation
\usepackage{soul}
\usepackage{tabularx}
\usepackage[caption=false]{subfig}

\begin{document}
\title{Distributed quantum sensing with a mode-entangled network of spin-squeezed atomic states} 
\author{Benjamin K. Malia}
\affiliation{Department of Physics, Stanford University, Stanford, CA, USA}
\affiliation{School of Applied and Engineering Physics, Cornell University, Ithaca, NY, USA}
\author{Yunfan Wu}
\affiliation{Department of Applied Physics, Stanford University, Stanford, CA, USA}
\author{Juli\'{a}n Mart\'{i}nez-Rinc\'{o}n}
\affiliation{Department of Physics, Stanford University, Stanford, CA, USA}
\affiliation{Quantum Information Science and Technology Laboratory, Instrumentation Division, Brookhaven National Laboratory, Upton, NY, USA}
\author{Mark A. Kasevich}
\email[]{kasevich@stanford.edu}
\affiliation{Department of Physics, Stanford University, Stanford, CA, USA}
\affiliation{Department of Applied Physics, Stanford University, Stanford, CA, USA}
\date{\today}

\begin{abstract}
Quantum sensors are used for precision timekeeping, field sensing, and quantum communication~\cite{Grotti2018,McGrew2018,Guo2020}. Comparisons among a distributed network of these sensors are capable of, for example, synchronizing clocks at different locations~\cite{Zhao2021,Zhang2021,Giovannetti2001,Beloy2021,Bothwell2022}. The performance of a sensor network is limited by technical challenges as well as the inherent noise associated with the quantum states used to realize the network~\cite{Pedrozo2020}. For networks with only local entanglement at each node, the noise performance of the network improves at best with square root of the number of nodes~\cite{Zheng2022}. Here, we demonstrate that nonlocal entanglement between network nodes offers better scaling with network size. A shared quantum nondemolition measurement entangles a clock network with up to four nodes. This network provides up to 4.5 dB better precision than one without nonlocal entanglement, and 11.6 dB improvement as compared to a network of sensors operating at the quantum projection noise limit. We demonstrate the generality of the approach with atomic clock  and atomic interferometer protocols, in scientific and technologically relevant configurations optimized for intrinsically differential comparisons of sensor outputs.
\end{abstract}

\pacs{}
\maketitle

Distributed quantum sensors detect and compare phase shifts between spatially distinct modes of quantum systems with high precision~\cite{Zhao2021,Zhang2021,Giovannetti2001}. For example, the gravitational potential can induce relative phase shifts between spatially separated atomic clocks \cite{Grotti2018} or atom interferometers~\cite{Overstreet2022}. Quantum systems are an attractive platform for networks as they have the unique ability to directly benefit from both local and nonlocal entanglement. Experiments have demonstrated entanglement-enhanced networks in both discrete~\cite{Liu2021} and continuous variable~\cite{Xia2020} configurations. In general, quantum networks will play an important role in future technologies. Significant progress has been made with networks of quantum systems~\cite{Lu2019,McGrew2018,Bodine2020,Beloy2021,Matsukevich2006,Chou2005,Simon2007} for enhanced communication~\cite{Muralidharan2016,Gundogan2021} and timekeeping~\cite{Komar2014,Polzik2016} at the global scale.

At small length scales, optical atomic clocks have pushed precision to record levels. In the work of Zheng et al.~\cite{Zheng2022}, up to six multiplexed Sr atomic clocks spaced over 1 cm are implemented to achieve a fractional frequency precision at the $10^{-20}$ level. Another work by Bothwell et al.~\cite{Bothwell2022} has measured the gravitational redshift over 1 mm within a single, spatially distributed sample of atomic Sr. In these systems, each clock's precision is limited by the Quantum Projection Noise (QPN) limit. In these mode-separable (MS) systems, the absence of correlation between the modes causes the total precision to scale as $1/\sqrt{M}$ where M is the number of identical clocks being compared. 

Through entanglement, a spin-squeezed clock or sensor is able to achieve precision beyond the QPN limit~\cite{Leroux2010a,Pedrozo2020}. However, if a network of squeezed clocks is MS, then the total precision still scales as $1/\sqrt{M}$. If nonlocal entanglement does exist, then the total precision of such a mode-entangled (ME) system has the potential to scale with the Heisenberg limit, 1/M~\cite{Komar2014,Polzik2016,Gessner2018,Zhuang2018,Eckert2006}. Guo et al. have demonstrated this scaling in a photonic system~\cite{Guo2020}, and Nichol et al. have measured this in a system of two Sr${}^+$ ions connected by a photonic link~\cite{Nichol2021}. Our work addresses a spin-squeezed ${}^{87}$Rb ME network whose noise scales better than a MS network.

\begin{figure*}[t]
\centering
\includegraphics[width=\linewidth]{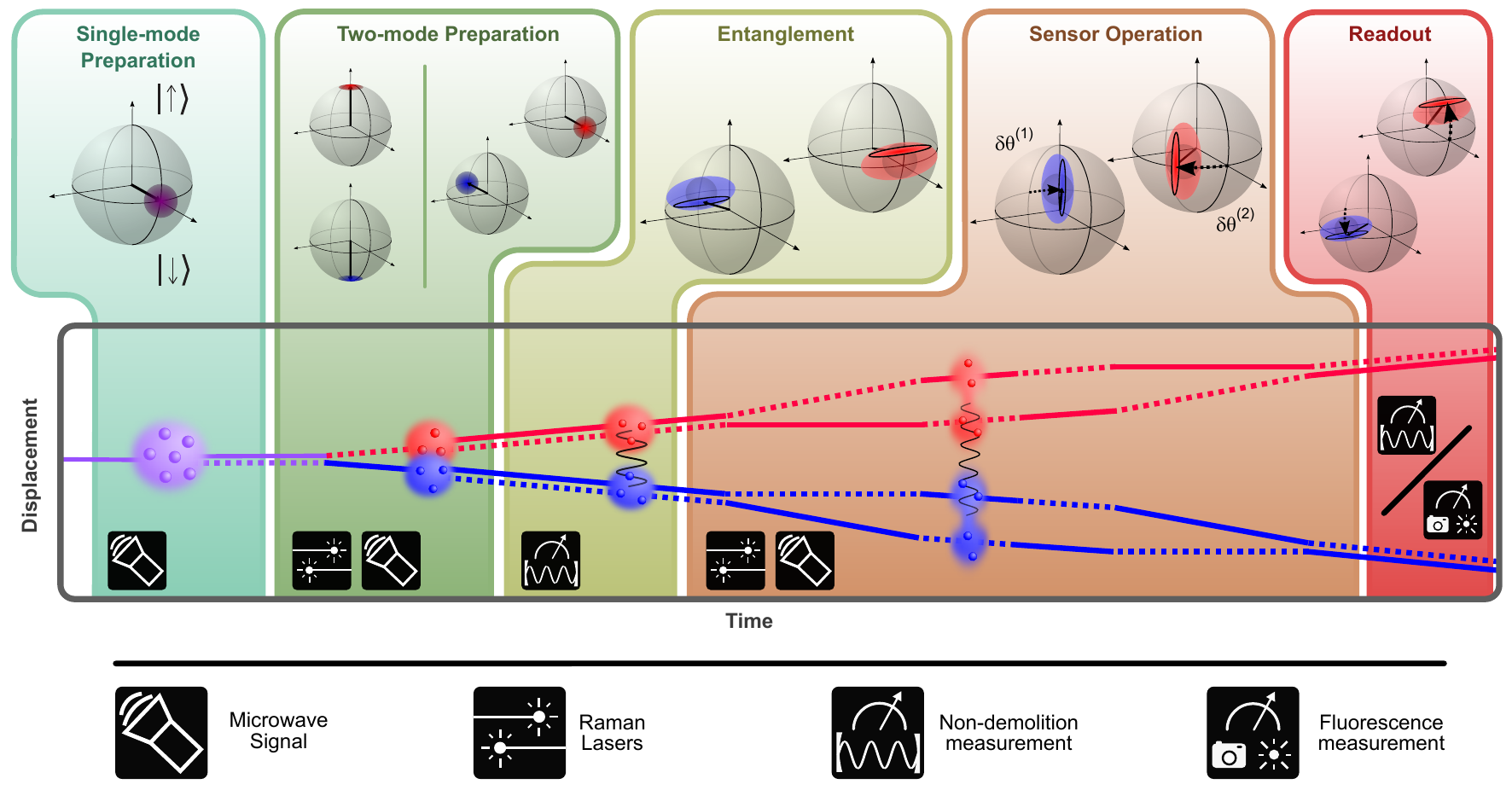}
\caption{\textbf{Atomic sensor sequence}. \textbf{(Single-mode preparation)} A localized ensemble of atoms (purple circles) are prepared in a $J_z=0$ CSS. The purple distribution on the Bloch sphere is the CSS's Wigner function with a variance of $N/4$. \textbf{(Two-mode preparation)} Counter-propagating Raman lasers split the ensemble into two spatially distinct modes (red and blue circles). Each mode is located opposite each other on their respective Bloch spheres (red and blue distributions). Note that since the modes are separable here, the two distributions are not dependent on each other. At this stage, a $\pi/2$ microwave pulse brings both of these states to $J_z=0$. \textbf{(Entanglement)} A probe laser performs a QND measurement to create mode-entanglement. Measuring each mode independently does not give enhanced precision on the total measurement (gray shaded distribution represents the CSS of each mode). To show how simultaneous measurement improves precision, an example of marginal (light distributions) and conditional (black outlined distributions) Wigner functions~\cite{Jing2019} are shown on the Bloch spheres. Here, the red mode squeezed above the equator is conditional on the blue mode being found below the equator. \textbf{(Sensor operation)} The sensor requires an initial application of a $\pi/2$ microwave pulse which rotates the SSS to a vertical (phase sensitive) orientation on the Bloch sphere. The observable being measured dictates the series of microwave and Raman pulses applied during the sensor sequence. The atomic interferometer sequence is pictured here, with a detailed description of its sequence described in Fig.~\ref{fig:pulsesAI} in Methods. Mean trajectories of spin down (up) states are represented by solid (dotted) lines. (Relative times are not to scale.) In the presence of a field gradient, the phases of the modes shift by $\delta\theta^{(m)}$ (dashed arrows). \textbf{(Readout)} A $\pi/2$ microwave pulse then rotates the states back to a horizontal orientation and a second measurement (either QND or fluorescence population spectroscopy) is performed to measure the shift in the sum of all spin values.} 
\label{fig:sequence}
\end{figure*}

Several methods exist for generating spin-squeezing between spatially separate modes. In the pioneering work of Julsgaard et al.~\cite{Julsgaard2001}, two spatially separated Rb vapor cells were probed via a photonic quantum non-demolition (QND) measurement. In Bose Einstein Condensates (BECs), on the other hand, spin-squeezing can be generated through local collisions before the state is allowed to expand to several micrometers~\cite{Fadel2018,Lange2018,Kunkel2018}. Each part of the cloud can then be imaged separately. More recently, Anders et al. separated an entangled BEC state even further, to 80 $\mu$m, with the application of velocity-dependant Raman transitions~\cite{Anders2021}. Not only does the spin system now occupy separate spatial modes, but the modes consist of states with differing momenta. Finally, atom-cavity interactions can entangle two momentum states with different spin states~\cite{Greve2021}.

In this work, we demonstrate a spatially distributed multimode atomic clock network with noise below the QPN limit. Velocity-dependent Raman transitions create up to four modes, each separated from an adjacent mode by $\sim$ 20 $\mu$m, before a nonlocal QND measurement is performed to entangle the modes. Nonlocal entanglement-enhanced precision is verified with networks of identical clocks, each containing 45,000 atoms per mode. The ME four-mode network exhibits noise roughly 4.5(0.8) dB lower than that of an equivalent MS network of spin-squeezed states (SSS) and 11.6(1.1) dB lower than a network of coherent spin states (CSS) operating at the QPN limit. Finally, we employ an $M=2$ node network to demonstrate an entangled differential atom interferometer.

The methods and apparatus used to generate and detect SSS are detailed in Refs.~\cite{Hosten2016a,Malia2020}. In summary, ensembles of up to 220,000 ${}^{87}$Rb atoms are cooled to 25 $\mu$K and trapped in a 1,560 nm 1D lattice within a dual wavelength optical cavity. This cavity enables QND measurements via a 780 nm probe detuned from the $D_2$ transition. These projective measurements detect and squeeze the ensemble's collective spin, $J_z = (N_\uparrow-N_\downarrow)/2$, where $N_i$ are the populations of atoms in each state after the measurement. This spin-1/2 system is defined by the hyperfine ground states of ${}^{87}$Rb, $\ket{\downarrow} = 5^2S_{1/2}\ket{F=1,m_F=0}$ and $\ket{\uparrow}= 5^2S_{1/2}\ket{F=2,m_F=0}$. 

\begin{figure*}[t]
\centering
\includegraphics[width=\linewidth]{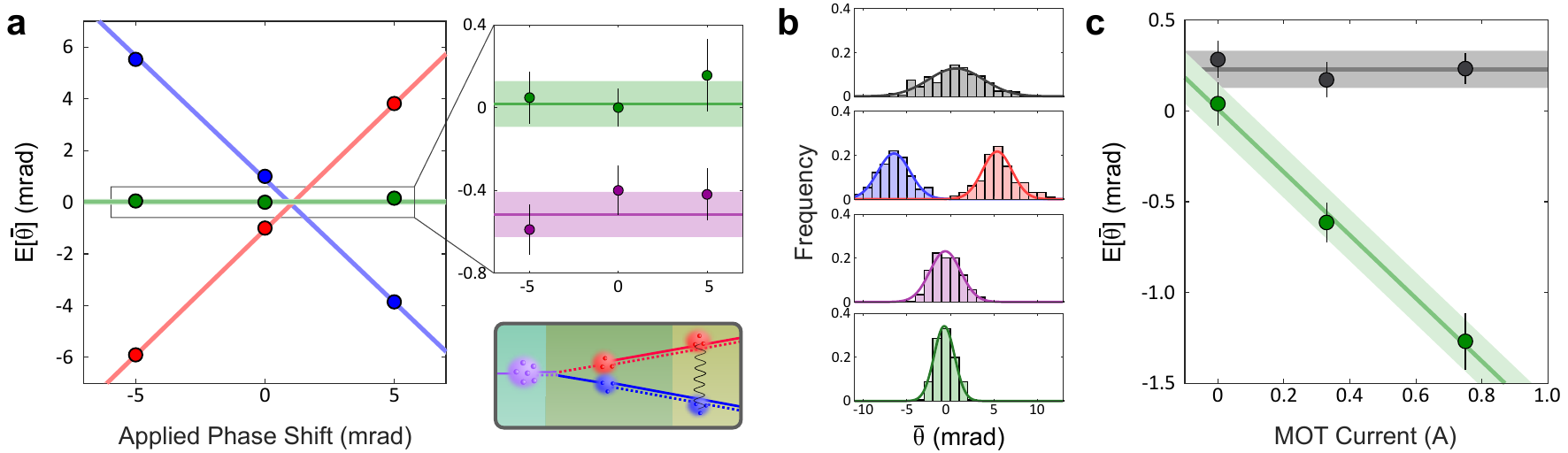}
\caption{\textbf{Differential phase shift detection.} \textbf{(a)} For the two-mode case depicted in the space-time diagram, the expected value of $\bar{\theta}$ is measured after the final microwave pulse is phase shifted. Solid lines are linear fits to the expected values for the positive momentum mode (blue), negative mode (red), and ME (green) cases. $\bar{\theta}$ in the single-mode cases are offset by 1 and -1 mrad respectively for visual clarity. The enlarged region contains the average of the single momentum modes, ie. MS states, (purple) which are offset by -0.4 mrad for visual clarity. In all subfigures, error bars represent the standard error of the mean (SEM) for a set of 200 samples and  shaded areas represent 68\% confidence intervals of the fits. \textbf{(b)} Distributions of 200 sample measurements for the two-mode sensor with coherent states (black), single-mode states (blue and red), MS states (purple), and ME states (green). Corresponding curves are Gaussian fits. \textbf{(c)} Response of a two-mode, ME sensor to a magnetic field gradient applied in the second half of the echo sequence (green circles). For reference, when the sensor sequence's microwave pulses are not performed (black circles), there is negligible change in $\bar{\theta}$ as the applied field increased. The relative magnetic field strength was determined by the relative voltage applied to the MOT coils.}
\label{fig:shift}
\end{figure*}

In order to generate spatially separate modes, velocity-dependent stimulated Raman transitions couple these spin states to momentum $\mathbf{p}$, where eigenstates are denoted as $\ket{\text{spin},\mathbf{p}}$. The relevant transitions are driven by $\pi$ pulses that take $\ket{\downarrow,\mathbf{p}_I} \rightarrow \ket{\uparrow,\mathbf{p}_I+2\hbar \mathbf{k}}$ and $\ket{\uparrow,\mathbf{p}_I} \rightarrow \ket{\downarrow,\mathbf{p}_I-2\hbar \mathbf{k}}$, where $\mathbf{k}$ is the effective wavevector associated with the Raman transition. Without loss of generality, the initial momentum $\mathbf{p}_I$ can be set to zero. A laser system drives Raman transitions between groundstate hyperfine levels (see Methods). The $\pi$ pulse time is short enough to address nearly the entire velocity distribution of the atom source. The transitions occur with a Rabi frequency of $\Omega_R = 2\pi\times 500$ kHz and the maximum transition probability for a single Raman $\pi$ pulse is 88\%.

When a spin state in an equal superposition of $\ket{\downarrow,0}$ and $\ket{\uparrow,0}$ experiences a Raman $\pi$ pulse, it coherently evolves into a linear superposition of the two modes $\ket{\uparrow,+2\hbar \mathbf{k}}$ and $\ket{\downarrow,-2\hbar \mathbf{k}}$ (see Fig.~\ref{fig:sequence}). To determine the coherence between the two modes, we apply a second Raman $\pi$ pulse a time $T$ after the first Raman pulse such that the states have drifted apart by a distance $v_\text{rel}T$ where $v_\text{rel} = 4 \hbar |\mathbf{k}|/m_{\text{Rb}} = 2.4$ cm/s is the relative velocity induced by the stimulated Raman interaction and $m_{\text{Rb}}$ is the mass of an atom. A final microwave $\pi/2$ pulse is then used to probe the coherence between the two modes (the microwave Rabi frequency is $\sim~2\pi \times 3$ kHz here and in the work described below). As $T$ increases, the coherence is observed to decay as $e^{-T/\beta}$ with a time constant $\beta=0.46~\mu$s due to the velocity spread ($\sim$~6.9 cm/sec) of the atomic source (see Methods). After roughly $T = 1.5~\mu$s (36 nm of separation) the contrast becomes negligible, indicating mode separation.

If no effort is made to coherently recombine these modes, the system can now be treated as a two-mode quantum network, where each mode $m$ has a collective spin length of $\big<J^{(m)}_x\big> = CN^{(m)}/2$ and a QPN limited variance ($\Delta J_z^{(m)})^2 = N^{(m)}/4$. Modes with nearly equal mean atom number $N$ can each be represented on composite Bloch spheres with radii $CN/2$, where the contrast $C = 78(3)\%$ is determined by fluorescence imaging. Spin-squeezing improves the measurement of a linear combination of the polar angle shifts $\delta\theta^{(m)}=\delta J^{(m)}_z/(C N/2)$, where $\delta J_z^{(m)}$ are the differences in spin values between a first and second measurement (see Fig.~\ref{fig:sequence}). In the remainder of this work the measurable quantity $\bar{\theta}$, determined from the shift in the collective $\delta J_z = \sum \delta J_z^{(m)}$ value, will refer to the mean of the angles
\begin{equation}
    \bar{\theta}  = \frac{1}{M}\sum^M_{m=1}\delta \theta^{(m)}= \frac{1}{CMN/2}\sum^M_{m=1}\delta J^{(m)}_z,
\end{equation}
and $\Delta\bar{\theta}$ to the square root of its variance. The second observation of the collective $J_z$ is accomplished with a second cavity QND measurement in the case of a clock network demonstration (following Ref~\cite{Hosten2016a}) or a precision fluorescence measurement in the case of an atom interferometer demonstration (following Ref.~\cite{Malia2020}).

To first demonstrate the effect of phase shifts on each mode, ensembles of 80,000 atoms are prepared in three different initial states: $\ket{\uparrow,2\hbar \mathbf{k}}$, $\ket{\downarrow,-2\hbar \mathbf{k}}$, or a superposition of the two. In the superposition case, waiting 0.9 ms separates the modes by roughly 20 $\mu$m, a significantly greater distance than the 36 nm coherence length identified above. The two modes' spins now point towards opposite poles on their respective Bloch spheres. As shown in Fig.~\ref{fig:sequence}, a $\pi/2$ microwave pulse is then applied to the atoms. The pulse brings their vectors to the equator of their Bloch spheres (with radius 20,000 in the third case). The mode with positive momentum, for example, is now in a superposition of $\ket{\downarrow,2\hbar \mathbf{k}}$ and $\ket{\uparrow,2\hbar\mathbf{k}}$. Since the microwave $\pi/2$ pulse simultaneously addresses both modes, the Bloch vectors remain anti-parallel. Finally, a (now nonlocal) QND measurement is performed to projectively squeeze the distributed states.  This operation leads to a nonlocal correlation of $J_z$ values between the modes while increasing the variance of the spin distributions in the x-y plane, as illustrated in Fig.~\ref{fig:sequence}.

Once the ME state has been prepared, a microwave $\pi/2-\pi-\pi/2$ spin echo sequence with $T_{\text{int}}=110~ \mu$s between each pulse is performed. The phases of the microwave pulses are adjusted to accomodate the AC Stark shift ($\sim$ 1 rad) induced by the entangling QND pulse so that the $J_z$ distributions are in metrologically sensitive configurations, as illustrated in Fig.~\ref{fig:sequence}. This is accomplished through observation of $J_z$ for the independently prepared modes $\ket{\uparrow,2\hbar }$ and $\ket{\downarrow,2\hbar \mathbf{k}}$. A second QND measurement determines the phase shift applied to the last microwave pulse. The two single-mode cases, $\bar{\theta}=\delta\theta^{(m)}$, experience nearly equal and opposite responses due to their anti-parallel spins (see Fig.~\ref{fig:shift}a). $\bar{\theta}$ in the ME case is consistent with the mean of the single-mode cases, indicating that each mode reacts oppositely to the applied shift. Therefore, a sensor utilizing this method  will suppress phase noise associated with the pulses used for coherent spin manipulation.  This property is useful for suppressing local oscillator noise in clock comparisons and optical phase noise in light-pulse atom interferometry applications (as demonstrated below).

 On the other hand, this type of sensor will measure a differential phase shift between the two modes due to, for example, position dependent fields~\cite{Fadel2022}. We demonstrate a non-zero differential measurement via the application of a magnetic field gradient across the 20 $\mu$m separation between the two modes. To introduce a clock frequency imbalance between the two modes, the magnetic field coils of the magneto-optical trap (MOT) are pulsed on during the second half of the echo sequence. As the magnetic field gradient (determined by the MOT coil current) increases, $\bar{\theta}$ is observed to shift away from zero (see Fig.~\ref{fig:shift}c). The measured shift of 1.7(0.3) mrad/A corresponds to an average clock frequency shift of $\delta\omega = \bar{\theta}/T_{\text{int}} = 2\pi\times15.7(2.8)$ Hz/A. Second order Zeeman shifts of this magnitude require 4.0(0.8) G/cm/A while the ${}^{87}$Rb atoms are in the presence of the 600 mG bias field. This value is consistent with the gradient estimated from the geometry of the MOT coils. We observed no significant increase in the width of the detection histograms (as shown in Fig.~\ref{fig:shift}b) for the relatively small ($\sim$ 1 mrad) differential phase shifts used in this work. These data demonstrate that this protocol can be used to measure the frequency difference between two distant entangled clocks through the observed (conditional) shift in $J_z$. Additional modes can be added, as described below, to detect higher order spatial correlations between modes.

Next, we evaluate the noise performance of entangled, multimode clock networks with $N=45,000$ atoms per mode. The metrological improvement, relative to their respective M-mode network of N-atom coherent states, can be quantified by a parameter $\xi_\text{net}^2$ derived from the generalized version~\cite{Gessner2020} of the Wineland squeezing parameter~\cite{Wineland1994}. For example, when M = 2, 
\begin{equation}
\xi_\text{net}^2 \equiv \frac{1}{C^2}\frac{\text{Var}\big(\delta J^{(1)}_z+\delta J^{(2)}_z\big)}{\text{Var}\big(\delta J^{(1),\text{CSS}}_z\big)+\text{Var}\big(\delta J^{(2),\text{CSS}}_z\big)},
    \label{eq:sqzparam}
\end{equation}
where the CSS variances are $N/4$ (see Methods). This parameter accounts for both the local and nonlocal entanglement since $\text{Var}\big(\delta J^{(2)}_z+\delta J^{(1)}_z\big)$ is the sum of both individual variances and the covariance between the two modes.

\begin{figure}[t]
\centering
\includegraphics[width=\linewidth]{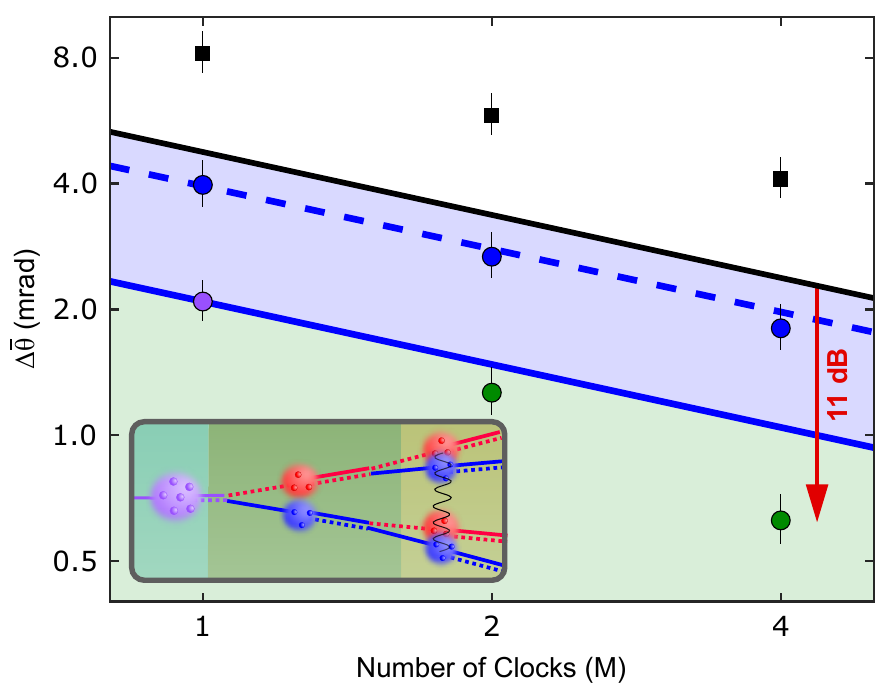}
\caption{\textbf{Clock network sensitivity}. Measured sensitivity for a clock network utilizing SSS. The QPN limit for $N = 45,000$ atoms is given by the black line, which is proportional to $1/\sqrt{M}$. A MS network is realized by taking independent measurements of a single-mode clock for each $\delta\theta^{(m)}$ component of $\bar{\theta}$. When observing the second QND measurement only (black squares) the sensitivity is above the QPN limit due to the QPN and local oscillator noise. The difference in QND measurements (blue circles) detects the spin-squeezing and brings the sensitivity below the QPN, however, the local oscillator noise remains. A one parameter fit to these data (dashed blue line) is consistent with $1/\sqrt{M}$ scaling. Reducing the local oscillator noise would push the sensitivity lower in the blue shaded area, where the lower limit (solid blue line) is determined by the $M=1$ sensitivity measured without a Ramsey sequence (purple point). The ME networks (green circles) exist below this limit, where the green area represents sensitivities obtainable exclusively through simultaneous measurements of a ME network. Error bars represent the pooled variance of three sets of 200 measurements. \textbf{(Inset)} Space-time diagram of the state preparation for a four-mode ME network.}
\label{fig:scaling}
\end{figure}

A two-mode SSS is prepared and a pair of $\pi/2$ microwave pulses separated by a time $T_{\text{int}}=100~\mu$sec constitute a standard Ramsey sequence. Squeezing reduces the variance of the joint measurement to $\Delta \bar{\theta} = 1.3(0.1)$ mrad (as shown in Fig.~\ref{fig:scaling}), which corresponds to $\xi_\text{net}^2=-8.6(1.0)$ dB. This precision is near that of a two-mode SSS in the absence of the Ramsey sequence ($\Delta \bar{\theta}=1.2(0.1)$ mrad without technical noise from the sensor sequence). A single-mode clock, on the other hand, has 3.6(0.6) mrad of technical noise. In this differential clock configuration, low measurement variance can be achieved without the need for high performance local oscillators, thus circumventing a limit of previous SSS sensor demonstrations~\cite{Hosten2016a}. This configuration will also suppress environmental noise common to both modes, such as a time varying bias magnetic field. This suppression is achieved with a single collective read-out measurement.

This method can be extended to $M=2^P$ clocks by further dividing the atomic ensemble into smaller subsets with $P$ Raman $\pi$ pulses. For example, we demonstrate a four-mode system by inserting an additional Raman $\pi$ pulse, followed by a microwave $\pi/2$ pulse, before the first QND measurement to generate spatially distinct modes. In this case, we adjust the total initial number of atoms to maintain $N=45,000$ atoms per mode. With four modes, the metrological enhancement is $\xi^2_\text{net} = 11.6(1.1)$ dB (see Methods). For comparison, this is a 4.5(0.8) dB relative improvement over the projected MS limit (see Fig.~\ref{fig:scaling}). Here, the network gain is driven by the improved squeezing efficiency for larger numbers of atoms since the total number of atoms initially entangled is $MN$.  The observed network gain is consistent with the measured atom number dependence of squeezing efficacy observed in Ref.~\cite{Hosten2016a} for this system. This four-mode network could be used to search, for example, for spatially periodic clock frequency shifts.

\begin{figure}[t]
\centering
\includegraphics[width=\linewidth]{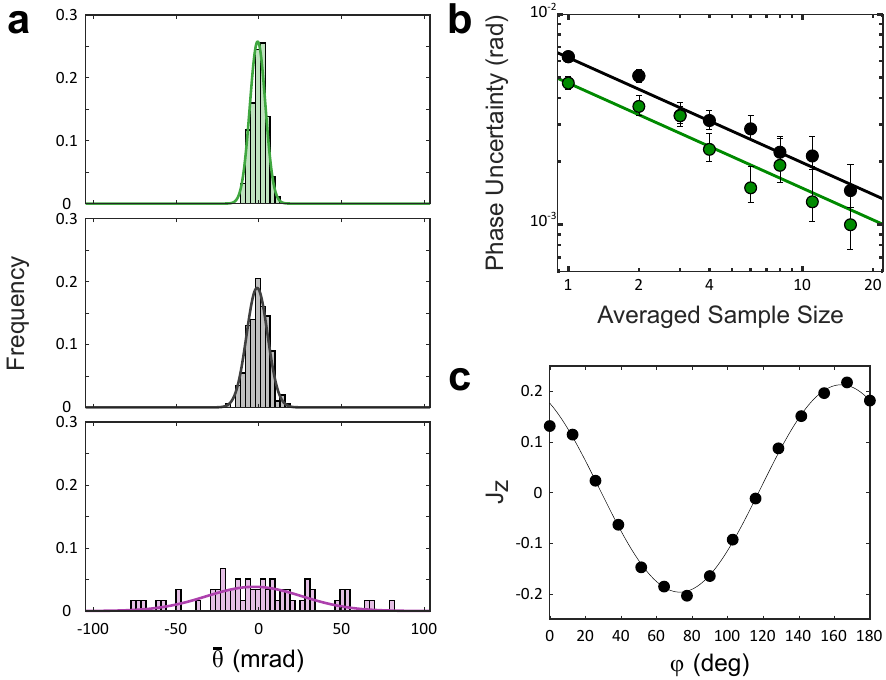}
\caption{\textbf{Interferometer performance}. \textbf{(a)} Distributions of 200 sample measurements for the two-mode sequence with ME squeezed states (green), MS coherent states (black), and for the single-mode sequence (purple). \textbf{(b)} The fractional stability for a two-mode interferometer with ME squeezed states (green) and MS coherent states (black) is calculated from a single data set of 200 samples. Error bars represent 68\% confidence intervals. \textbf{(c)} Response of a single-mode interferometer to a phase shift $\phi$ in the final Raman $\pi$ pulse of the sequence. The coherence is determined from the peak-to-peak value of a sinusoidal fit.}
\label{fig:intHist}
\end{figure}

Finally, we apply this method to an atom interferometer configuration, as illustrated in Figs.~\ref{fig:sequence} and ~\ref{fig:pulsesAI}. In this case, two atomic modes are initially entangled as described above in an optical lattice. Atoms are then released from the lattice and subject to an atom interferometer pulse sequence after an interval of approximately 7 msec, after which they have separated by $\sim$ 0.16 mm. Specifically, a Raman $\pi$ pulse acts as a beamsplitter by simultaneously imparting opposite momentum to the spin states in each branch (resulting in a relative momentum between interfering wavepackets of $4 \hbar \mathbf{k}$, as depicted in Fig.~\ref{fig:sequence} and~\ref{fig:pulsesAI}). $T_{\text{int}}=50~\mu$s later, a sequence of a Raman $\pi$ pulse, microwave $\pi$ pulse and Raman $\pi$ pulse act as a mirror, while a final Raman $\pi$ pulse recombines the states. The duration of the interferometer pulse sequence is 270 $\mu$sec, dominated by the $\sim$ 160 $\mu$sec microwave $\pi$ pulse time. Each mode of N = 110,000 atoms accumulates a phase proportional to its local acceleration (see Methods). Fluorescence imaging (Ref.~\cite{Malia2020}) detects of the sum of the final $J_z^{(m)}$ (the modes are too closely spaced to resolve individually on the camera). This differential method suppresses large common mode optical phase fluctuations associated with the optical stimulated Raman transitions (measured to be 10 mrad, or roughly 15 dB above the projection noise limit, see Fig.~\ref{fig:intHist}a).

The smallest observed single-shot phase uncertainty with a ME interferometer is 4.9(0.4) mrad (Fig.~\ref{fig:intHist}b), which corresponds to an inferred differential acceleration sensitivity of $1.4(0.1)\times10^{-2}~\text{m/sec}^2$ (see Methods). This sensitivity is limited by the relatively poor contrast (40\%, see Fig.~\ref{fig:intHist}c) associated with the interferometer pulse sequence. Entanglement-enhanced noise performance can be characterized by comparing the observed ME sensor noise to the noise observed for the same sensor sequence implemented without the entangling probe, as shown in Figs.~\ref{fig:intHist}a,b. With respect to the sequence which does not employ entanglement, we observe an average metrological improvement of 1.6(0.9) dB. The absolute noise is 0.1(0.7) dB above the QPN limit for the non-entangled sensor which is likely due to imperfect suppression of Raman laser phase noise. This configuration extrapolates directly to high performance, single source, differential gravity sensors (for example, Ref~\cite{Overstreet2022}). 

In the future, a distributed array of cavities sharing a common QND measurement~\cite{Polzik2016}, possibly via photonic links and shared probe light~\cite{Nichol2021,Chaudhary2022}, would enable entanglement across longer distances. Adapting this method to squeezed optical clocks~\cite{Pedrozo2020} would further push the limits of precision measurements of time~\cite{Beloy2021,Zheng2022} and gravity~\cite{Bothwell2022}. Applications in secure time transfer and quantum communications can benefit from a distributed entangled state~\cite{Komar2014} since an eavesdropper could not deduce the correlations through observation of one clock alone. For example, information encoded by rotations on one network node would only be detectable through a collective measurement of all nodes. Finally, the atomic interferometer protocol is technologically useful for future high performance gravity gradient sensors and differential configurations designed for gravitational wave detection~\cite{Abe2021,Zhan2019} and dark matter searches~\cite{Wcislo2018,Safronova2018,Tino2021}.

\begin{acknowledgments}
We acknowledge support from Department of Energy award DE-SC0019174-0001, the Department of Energy Q-NEXT NQI, a Vannevar Bush Faculty Fellowship and NSF QLCI Award OMA-2016244.

B.K.M., Y.W., and J.M. designed, constructed, and characterized the experiment. B.K.M. and Y.W. performed data collection and analysis. M.A.K. supervised the research. All authors contributed to the manuscript.
\end{acknowledgments}

\section{Methods}

\subsection{Contrast}
\begin{figure}[t]
\centering
\includegraphics[width=\linewidth]{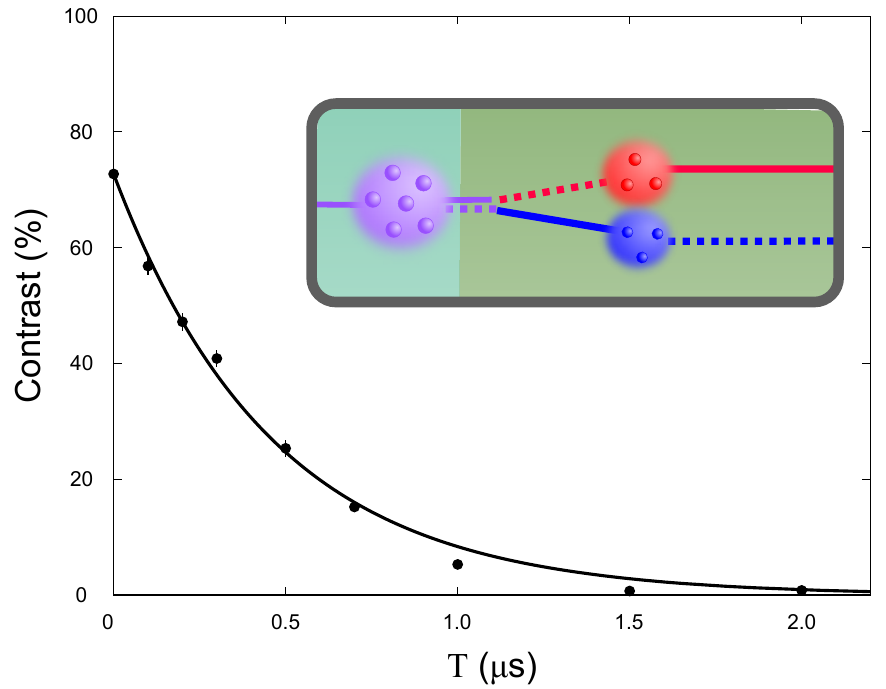}
\caption{\textbf{Mode separation.} Contrast of the collective fluorescent measurement as a function of separation time between two $0.33~\mu$s Raman $\pi$ pulses. Solid curve is an exponential fit to the data with a decay rate of 0.46 $\mu$s. Note that $T=0$ corresponds to a single pulse with a total time of $2\pi$. Error bars represent a 95\% confidence interval.}
\label{fig:contrast}
\end{figure}

When determining mode separation, a second Raman pulse removes the relative momentum after time $T$ to maintain the mode separation distance until detection takes place. A $\pi/2$ microwave pulse with varying phase addresses both modes simultaneously and the remaining contrast is determined by the peak-to-peak $J_z$ values from fluorescence imaging (see Fig.~\ref{fig:contrast}). The coherence falls to zero after roughly one thermal debroglie wavelength $\lambda_\text{th}=h/\sqrt{2\pi m_\text{Rb}k_BT_\text{ens}} = 36$ nm, where $k_B$ is the Boltzmann constant, $T_\text{ens}$ is the temperature of the ensemble, and $m_\text{Rb}$ is the mass of the ${}^{87}$Rb atom. In the absence of Raman transitions, the contrast is $C = 79(1)\%$ due to decoherence in the lattice both before and after the sensing times. Adding in the two Raman transitions with $T=0$ decreases the contrast to  $C = 73(1)\%$.

To determine $C$ for the clock measurement, a final microwave $\pi/2$ pulse temporarily introduced to the single-mode case resolves $C = 78(3)\%$. Therefore, introducing a single Raman transition before the QND measurement does not significantly reduce the final coherence of the ensemble. The gravity gradiometer has a lower final coherence, roughly 40\%, due to four additional Raman pulses. This is consistent with the expected $C = (88\%~\text{population transfer})^4\times(79\%~\text{contrast without gradiometer})$.

\subsection{Squeezing matrix for multiparameter discrete-variable squeezing}

The form of the metrological squeezing parameter $\xi_\text{net}^2$ is derived from Equation 13 in Gessner et al.~\cite{Gessner2020}. For a general multimode system, the squeezing matrix $\Xi^2$ characterizes the level of metrological improvement due to entangled quantum network. In other words, it compares the covariances between each mode to the QPN limit. The matrix elements can be defined as
\begin{equation}
\Xi^2_{kl}= \frac{\sqrt{N^{(k)} N^{(l)}} \text{Cov}\big(\hat{J}^{(k)}_z,\hat{J}^{(l)}_z\big)}{\big<\hat{J}^{(k)}_x\big>\big<\hat{J}^{(l)}_x\big>},
\end{equation}
where $\hat{J}^{(m)}_x$ are the spin operators for each mode and the mean spins are in the $\mathbf{\hat{x}}$ direction. 

The metrological improvement in the multiparameter estimation can be written as the ratio of variance of the squeezed network to that of a network comprised of coherent states:  $\mathbf{n}^T\mathbf{\Sigma}^2\mathbf{n}/\mathbf{n}^T\mathbf{\Sigma_\text{SN}}^2\mathbf{n}$, where $\mathbf{\Sigma}$ and $\mathbf{\Sigma_\text{SN}}$ are the covariance matrices of the squeezed and coherent states respectively, and $\mathbf{n}$ is the vector of coefficients for the linear combination of parameters being measured. In the case of equally populated [$N = N^{(m)}$] modes, the expected length values are $\big<\hat{J}^{(m)}_x\big>=CN/2$. For a measurement of the average angular shift ($n_m = 1/M$), it can be shown that the metrological improvement reduces to $\xi_\text{net}^2=M\mathbf{n}^T\mathbf{\Xi^2}\mathbf{n}$. More explicitly, in terms of the measured observables, it can be written as
\begin{equation}
\xi_\text{net}^2 \equiv \frac{1}{C^2}\frac{\text{Var}\big(\sum^M_{m=1}\delta J^{(m)}_z\big)}{M\times\text{Var}\big(\delta J^{\text{CSS}}_z\big)}.
    \label{eq:gensqzparam}
\end{equation}

\subsection{Measurement Sensitivity}

Since the QND measurement addresses all modes simultaneously, it cannot distinguish between spin states with different momenta. The measured $\delta J_z$ is simply the sum of $\delta J^{(m)}_z$, with expectation value
\begin{equation}
    \big<\delta J_z\big> = \sum_{m=1}^M \delta J^{(m)}_z = C\frac{N}{2}\sum_{m=1}^M \delta \theta^{(m)} =C\frac{N}{2}M\bar{\theta}.
\end{equation}
The sensitivity, $\sigma$, of this measurement to changes in $\bar{\theta}$ is given by standard error propagation~\cite{Guo2020}:
\begin{equation}
    \sigma = \frac{\sqrt{\text{Var}(\delta J_z)}}{\partial\big<\delta J_z\big>/\partial\bar{\theta}} = \frac{\Delta (\delta J_z)}{CMN/2}=\Delta\bar{\theta}.
\end{equation}

\begin{figure}[t!]
\centering
\includegraphics[width=\linewidth]{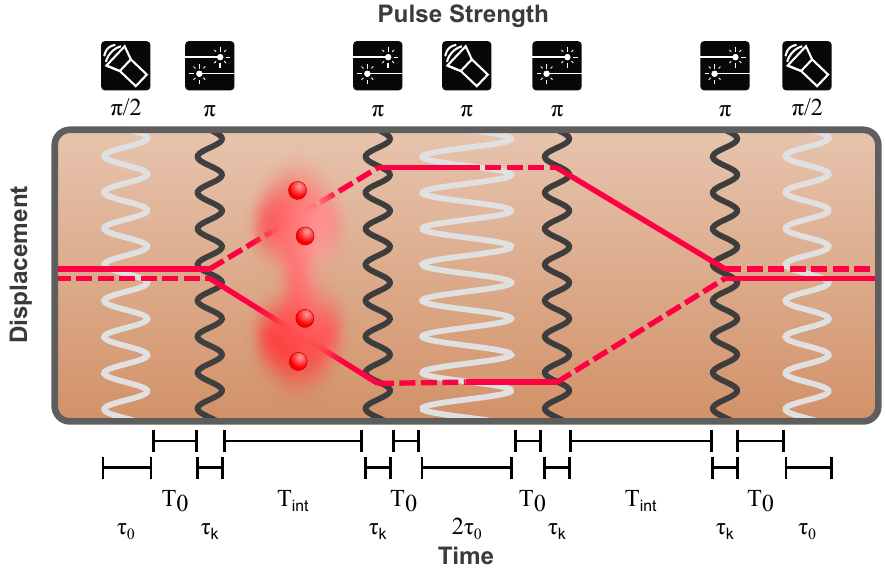}
\caption{\textbf{Interferometer sequence timing.} Space time diagram in the inertial frame of a single-mode interferometer. Solid (dashed) lines represent the trajectory of the spin down (up) state. White (gray) waves represent the finite time of the microwave (Raman) pulses.}
\label{fig:pulsesAI}
\end{figure}

\subsection{Laser System}

A low phase noise, 1,560 nm laser is frequency doubled to 780 nm. This light is split and one mode passes through an electro-optic modulator (EOM) driven at 6.434 GHz, 400 MHz lower than the hyperfine transition frequency, $\omega_{\text{HF}}$. The driving signal is created by a low phase noise crystal oscillator mixed with a direct digital synthesizer (DDS), which allows for power, frequency and phase control. Next, both modes are amplified by semiconductor-based optical amplifiers to 2.8 W each. 

One mode is now up-shifted by a 200 MHz acousto-optic modulator (AOM) and the other is down-shifted by the same amount. Both AOMs are driven by a common signal from a low noise 200 MHz crystal oscillator. The pulsed signal controls the time the AOMs couple the light to optical fibers which deliver the light to the atoms. The fibers launch the light into 5.4 mm diameter, counter-propagating freespace beams at 45 deg angle to the vertical and a 45 deg angle to the cavity axis. The shifting places one sideband of the modulated beam $\omega_{\text{HF}}$ away from the un-modulated beam frequency. These two frequencies drive the Raman transition between the two hyperfine states. The two participating frequencies create a transition which is red-detuned by 3.5 GHz from the excited state. The other sidebands are used to balance the AC-Stark shift and do not contribute significantly to the population change.

\subsection{Interferometer Phase Shift}

The sequence provided in this work differs from a standard Mach-Zehnder configuration~\cite{Kasevich1991} in that both spin states receive a momentum kick instead of just one state. In addition, the microwave pulses are longer than the interrogation time so terms including pulse durations must be considered. The total phase shifts, $\delta \theta^{(m)}$, of an interferometer can be derived from the sensitivity function~\cite{Malia2021_thesis}:
\begin{equation}
\begin{split}
\delta \theta^{(m)} = 2|\mathbf{k}|a^{(m)}(2T_{\text{int}}^2 + 4 T_{\text{int}} T_0 + 4 T_{\text{int}}\tau_0 \\+ 6 T_{\text{int}} \tau_k + 4 T_0 \tau_k + 4 \tau_0 \tau_k + 4 \tau_k^2),
\end{split}    
\end{equation}
where $a^{(m)}$ is the acceleration in mode $m$ projected along $\mathbf{k}$, $T_0 = 1 ~\mu$sec is the time between sequential pulses, $\tau_0=80$ $\mu$sec is the duration of a microwave $\pi/2$ pulse, and $\tau_k=2~\mu$sec is the duration of a Raman $\pi$ pulse (see Fig.~\ref{fig:pulsesAI}). For the data of Fig.~\ref{fig:intHist}b, where $T_{\text{int}} = 50~\mu$sec, we infer a statistical sensitivity of $\Delta \bar{a} = (\sum^M_{m=1} a^{(m)})/M = 1.4(0.1)\times10^{-2}~\text{m/sec}^2$ for a single shot.

\end{document}